\documentclass[conference]{IEEEtran}
\IEEEoverridecommandlockouts

\usepackage{cite}
\usepackage{amsmath,amssymb,amsfonts}
\usepackage{algorithmic}
\usepackage{graphicx}
\usepackage{textcomp}
\usepackage[colorlinks]{hyperref}
\hypersetup{citecolor=blue}
\usepackage[table]{xcolor}
\usepackage{balance}
\usepackage{xcolor}
\usepackage{float}  

\def\BibTeX{{\rm B\kern-.05em{\sc i\kern-.025em b}\kern-.08em
    T\kern-.1667em\lower.7ex\hbox{E}\kern-.125emX}}
\begin{document}

\title{Neural Networks-Enabled Channel \\Reconstruction for Fluid Antenna Systems: \\A Data-Driven Approach}
\author{Haoyu Liang, Zhentian Zhang, Jian Dang, Hao Jiang, Zaichen Zhang
	\thanks{ }
	\thanks{Haoyu Liang, Zhentian Zhang, Zaichen Zhang are with the National Mobile Communications Research Laboratory, Frontiers Science Center for Mobile Information Communication and Security, Southeast University, Nanjing, 210096, China. Zaichen Zhang are also with the Purple Mountain Laboratories, Nanjing 211111, China. (e-mails: \{zhangzhentian, zczhang\}@seu.edu.cn).
	
	Jian Dang is with the National Mobile Communications Research Labo-ratory, Frontiers Science Center for Mobile Information Communication and Security, Southeast University, Nanjing 211189, China, also with the Key Laboratory of Intelligent Support Technology for Complex Environments, Ministry of Education, Nanjing University of Information Science and Tech-nology, Nanjing 210044, China, and also with Purple Mountain Laboratories, Nanjing 211111, China.(email: dangjian@seu.edu.cn). 
	
	Hao Jiang is with School of Artificial Intelligence, Nanjing University of Information Science and Technology, Nanjing 210044, China. (email: jianghao@nuist.edu.cn)}
		\thanks{This work of Jian Dang and Zaichen Zhang is partly supported by the Fundamental Research Funds for the Central Universities (2242022k60001), Basic Research Program of Jiangsu (No. BK20252003), the Key Laboratory of Intelligent Support Technology for Complex Environments, Ministry of Education, Nanjing University of Information Science and Technology (No. B2202402). The work of Hao Jiang is partly supported in part by the National Natural Science Foundation of China (NSFC) projects (No. 62471238).}
		}

\maketitle

\begin{abstract}
Fluid antenna systems (FASs) offer substantial spatial diversity by exploiting the electromagnetic port correlation within compact array spaces, thereby generating favorable small-scale fading conditions with beneficial channel gain envelope fluctuations. This unique capability opens new opportunities for a wide range of communication applications and emerging technologies. However, accurate channel state information (CSI) must be acquired before a fluid antenna can be effectively utilized. Although several efforts have been made toward channel reconstruction in FASs, a generally applicable solution to both model-based or model-free scenario with both high precision and efficient computational flow remains lacking. In this work, we propose a data-driven channel reconstruction approach enabled by neural networks. The proposed framework not only achieves significantly enhanced reconstruction accuracy but also requires substantially lower computational complexity compared with existing model-free methods. Numerical results further demonstrate the rapid convergence and robust reconstruction capability of the proposed scheme, outperforming current state-of-the-art techniques.
\end{abstract}

\section{Introduction}
\subsection{Background}
Fluid antenna systems (FASs) introduce a novel train of thought in re-configurable antenna–oriented wireless communication network designs \cite{fas_tur}, offering substantial spatial diversity and architectural simplicity. A fluid antenna refers to any software-controllable fluidic, conductive, or dielectric structure capable of dynamically altering its radiation characteristics according to system requirements \cite{fas_tur2, fbl_fas}. Different from the prevailing evolution of multi-input and multi-output (MIMO) systems, FASs exploit the natural coupling effect inherent in compact antenna spaces (an effect neglected by conventional MIMO systems with fixed antenna arrays) to generate significant spatial gains \cite{fas_tur3}, thereby mitigating the need for costly hardware expansion. FAS technology is reshaping contemporary network designs across multiple access \cite{fas_ma, fas_ma2}, channel modeling \cite{fas_modeling}, signal processing \cite{fas_signal}, modulation \cite{fas_modulation}, and emerging paradigms such as joint communication and sensing \cite{fas_isac, fas_isac2}.

\subsection{Related Work}
In practical FAS designs, the problem of reasonable port selection \cite{fas_port} ensures that only the most favorable channels are used in uplink or downlink transmission. To achieve this, channel state information (CSI) must first be estimated, after which an effective activation pattern for the fluid antenna ports can be selected. Consequently, CSI estimation or channel reconstruction becomes a central task and is also the main focus of this work. The following subsections summarize representative state-of-the-art approaches.

\subsubsection{Greedy-based Sparse Regression}
Under geometric transmission models with a finite number of scatterers, channel reconstruction in FAS can be formulated as a sparse linear regression problem \cite{greedy1,greedy2}, solvable via compressive sensing techniques. Greedy algorithms such as the matching pursuit (MP) family reconstruct sparse signals from limited measurements. Beginning with the observation vector as the initial residual and an empty support set, greedy algorithms iteratively identify the dictionary atom most correlated with the residual, append its index to the support set, and update both the sub-dictionary and residual using least-squares or minimum mean square error (MMSE) estimators \cite{ls_estimate}.

\subsubsection{Bayesian Learning-based Algorithms}
The aforementioned greedy approaches highly rely on prior knowledge of channel models. However, in practice, such channel models may be unknown. To address this, self-adaptive methods have been proposed. In \cite{fas_amp}, leveraging the law of large numbers, the statistical distribution of the unknown channel coefficients is learned and incorporated into the reconstruction process. Using a low-complexity approximate message passing (AMP) framework, estimation and distribution updates are performed via Bayesian inference.

Another adaptive algorithm, termed S-BAR \cite{fas_beyes}, models the port-domain channel as a stochastic process and exploits spatial correlation through an experiential kernel within a sampling–regression framework. Offline, it greedily selects informative ports and precomputes reconstruction weights. Online, fluid antennas sample via the designed switch matrix, enables full channel recovery from only a few pilots. However, S-BAR suffers from prohibitively large computational complexity on the order of cubic time.

\subsection{Contributions}
In this work, we propose a data-driven approach for channel reconstruction in FASs. Unlike the aforementioned state-of-the-art methods, our approach does not rely on specific signal assumptions such as sparsity structures (required by greedy algorithms) or predefined statistical distributions (required by Bayesian methods). Instead, we introduce a neural network–based reconstruction framework. Specifically, the proposed method processes the received pilot signals into a trainable dataset, which is then fed into a carefully designed neural network for channel reconstruction.

Compared with many existing methods, the proposed framework achieves significantly improved estimation accuracy. Notably, whereas model-free approaches such as S-BAR incur cubic computational complexity, the proposed method reduces complexity to between linear and quadratic order, while maintaining superior performance under model-free FAS conditions. 

{\em The reproducible simulation codes are available at https://github.com/BrooklynSEUPHD/Neural-Networks-Enabled-Channel-Reconstruction-for-Fluid-Antenna-Systems.git}

The remainder of the paper is organized as follows. Sec.~\ref{sec.system} describes the FAS system configuration and signal generation. Sec.~\ref{sec.proposed} details the proposed data-driven framework and the corresponding learning procedures. Sec.~\ref{sec.numerical} presents numerical results, including estimation performance and convergence behavior. Finally, conclusions are drawn in Sec.~\ref{sec.conclusion}.
\section{System Model}\label{sec.system}
\begin{figure}[!t]
	\centering
	\includegraphics[width=\columnwidth]{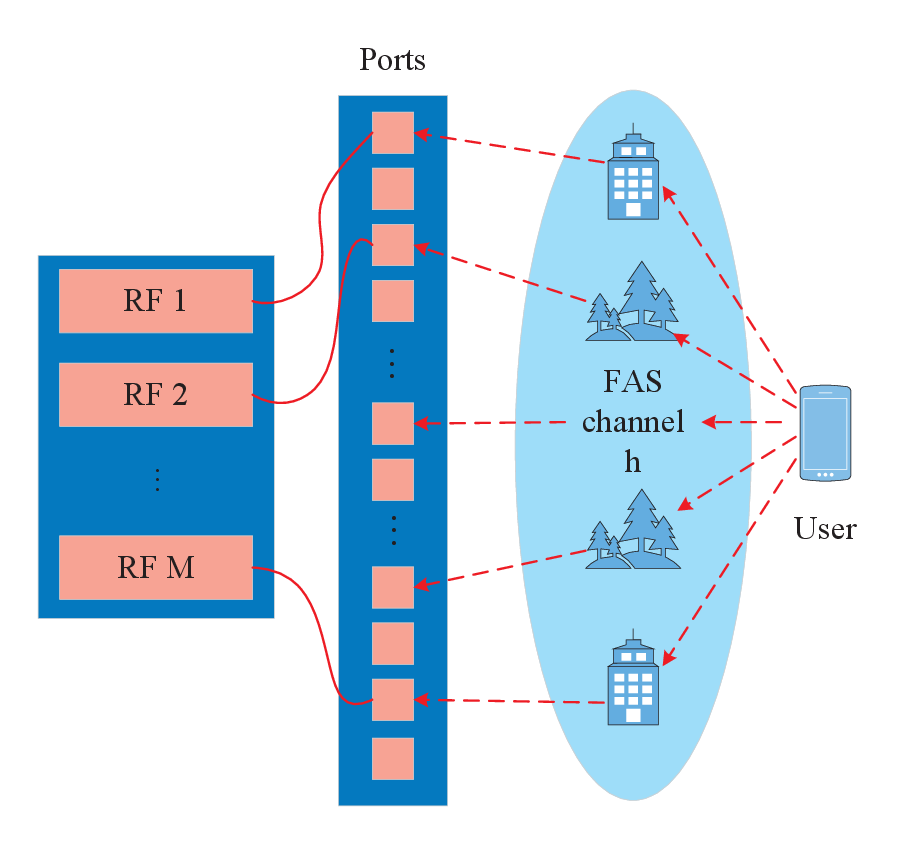}
	\caption{An illustration of channel estimation for an FAS.}
	\label{fig:CE_FAS}
\end{figure}

We consider an uplink FAS network consisting of $M$ fluid antennas, a base station (BS) equipped with $N$ ports, and a single-antenna user. As illustrated in Fig.~\ref{fig:CE_FAS}, the fluid antenna is arranged linearly over a length of $W\lambda$ and comprises $N$ uniformly spaced ports, where $\lambda$ denotes the wavelength. Each fluid antenna is connected to an RF chain dedicated to pilot reception, and its position can be switched to one of the $N$ available port locations.

For a point-to-point FAS, in which the transmitter employs a fixed antenna while the receiver is equipped with a fluid antenna, i.e., a single-input multiple-output (SIMO) configuration, the received signal at the $j$-th port slot of the fluid antenna can be expressed as follows:
\begin{equation}
	y_j = h_j x_p + z_j,
\end{equation}
where $h_j$ denotes the channel vector corresponding to the $j$-th port, $x_p$ is the pilot transmitted by the user, and $z_j \sim \mathcal{C N}(0, \sigma^2)$ is the additive white Gaussian noise (AWGN).

Define the switching matrix as the position of the fluid antenna within the port. The $(n,m)$-th entry of the matrix equals 1 if the $m$-th antenna is placed at port $n$, and 0 otherwise. Due to hardware constraints, exactly $M$ of the $N$ ports must be selected per timeslot. Accordingly, each column of $\mathbf{S}_p$ contains a single 1, and no two 1s appear in the same row, ensuring that $M$ distinct ports are chosen. The switching matrix at the $p$-th pilot slot is defined as $\mathbf{S}_p \in \mathcal{S}$, where:
\begin{equation}
	\begin{aligned}
		\mathcal{S}
		= & \Big\{ \mathbf{S}_p \in \{0,1\}^{N \times M} \mid \\
		& \sum_{n=1}^N [\mathbf{S}_p]_{n,m} = 1, \quad \sum_{m=1}^M [\mathbf{S}_p]_{n,m} \le 1 \Big\}.
	\end{aligned}
\end{equation}

Each column of $\mathbf{S}_p$ represents the port-selection vector of a fluid antenna, ensuring that each antenna connects to one port only, while each port can serve at most one antenna.

Thus, for all $p \in [P]$, the switching matrix satisfies:
\begin{equation}
	\mathbf{S}_p^{\mathrm{H}} \mathbf{S}_p = \mathbf{I}_M.
\end{equation}

At the $p$-th pilot slot, the received signal $\mathbf{y}_p$ of selected ports can be expressed as:
\begin{equation}
	\mathbf{y}_p = \mathbf{S}_p^{\mathrm{H}} \mathbf{h}_p x_p + \mathbf{z}_p.
\end{equation}

Without loss of generality, we assume $x_p = 1$ for all $p \in [P]$. Considering all $P$ timeslots within a coherence frame, the aggregate received signal can be written in compact form as:
\begin{equation}
	\mathbf{y} = \mathbf{S}^{\mathrm{H}} \mathbf{h} + \mathbf{z},
\end{equation}
where $\mathbf{y} := \left[\mathbf{y}_1^{\mathrm{H}}, \ldots, \mathbf{y}_P^{\mathrm{H}}\right]^{\mathrm{H}}$, $\mathbf{S} := \left[\mathbf{S}_1, \ldots, \mathbf{S}_P\right]$, and $\mathbf{z} := \left[\mathbf{z}_1^{\mathrm{H}}, \ldots, \mathbf{z}_P^{\mathrm{H}}\right]^{\mathrm{H}}$.

The objective is to reconstruct the $N$-dimensional channel vector $\mathbf{h}$ from the noisy observation $\mathbf{y} \in \mathbb{C}^{PM}$. To accomplish this, existing studies have predominantly relied on compressed sensing (CS) and/or alternating optimization (AO) based methods, which are briefly reviewed in the following subsection.

For channel realizations, the wireless propagation between the user and the BS is modeled as a clustered scattering channel (CSC), which captures the spatial characteristics of multipath propagation in rich scattering environments. The channel vector $\mathbf{h} \in \mathbb{C}^N$ associated with the $N$ receive ports is given by:
\begin{equation}
	\mathbf{h} = \sqrt{\frac{N}{C R}} \sum_{c=1}^C \sum_{r=1}^R g_{c, r} \mathbf{a}\left(\theta_{c, r}\right),
\end{equation}
where $C$ denotes the number of scattering clusters, and each cluster contributes $R$ propagation rays. The complex coefficient $g_{c, r} \sim \mathcal{C} \mathcal{N}(0,1)$ represents the small-scale fading gain of the $r$-th ray in the $c$-th cluster, while $\theta_{c, r}$ denotes its angle of arrival (AoA). The term $\mathbf{a}(\theta_{c, r}) \in \mathbb{C}^N$ is the array steering vector corresponding to the AoA $\theta_{c, r}$, which for a uniform linear array (ULA) can be written as:
\begin{equation}
	\mathbf{a}(\theta) = \frac{1}{\sqrt{N}} \left[1, e^{j 2 \pi \frac{d}{\lambda} \cos (\theta)}, \ldots, e^{j 2 \pi \frac{d}{\lambda}(N-1) \cos (\theta)}\right]^{\mathrm{T}},
\end{equation}
where $d$ is the inter-element spacing and $\lambda$ is the carrier wavelength.

The normalization factor $\sqrt{\frac{N}{C R}}$ ensures that the average channel power satisfies $\mathbb{E}\left[\|\mathbf{h}\|^2\right] = N$. This clustered model effectively captures both spatial correlation and angular dispersion, and has been widely adopted in the analysis of FAS and MIMO systems. It provides a flexible framework for simulating diverse propagation environments by adjusting the parameters $C$, $R$, and the angular distribution of $\theta_{c, r}$.

\section{Neural Network-Based\\ Channel Reconstruction}\label{sec.proposed}
In this section, we first present the multilayer perceptron (MLP)-based channel estimation method and subsequently analyze its online estimation complexity. Overall, the proposed algorithm follows a data-driven channel reconstruction paradigm consisting of the following steps:
\begin{itemize}
	\item[i)] Dataset loading and normalization.
	\item[ii)] Lightweight multilayer fully connected neural network (FCNN) modeling.
	\item[iii)] Normalized mean squared error (NMSE)-driven supervised learning with early stopping.
\end{itemize}
The simplicity of the fully connected architecture enables efficient training while providing strong nonlinear approximation capability for handling complex wireless channel estimation tasks.
\subsection{Network Architecture}

To reconstruct the CSC, a FCNN with two hidden layers is designed to learn the nonlinear mapping from noisy pilot observations to the underlying channel vector. Let $\mathbf{r} \in \mathbb{C}^{PM}$ denote the concatenated pilot signals received at the BS, and let $\hat{\mathbf{h}} \in \mathbb{C}^N$ denote the estimated channel vector. For training convenience, both the input and output are transformed into real-valued forms by concatenating their real and imaginary parts:
\begin{equation}
	\tilde{\mathbf{r}}=\left[\Re(\mathbf{r})^{\mathrm{T}}, \Im(\mathbf{r})^{\mathrm{T}}\right]^{\mathrm{T}}, 
	\quad 
	\tilde{\mathbf{h}}=\left[\Re(\mathbf{h})^{\mathrm{T}}, \Im(\mathbf{h})^{\mathrm{T}}\right]^{\mathrm{T}}.
\end{equation}

\begin{figure}[!t]
	\centerline{\includegraphics[width=\columnwidth]{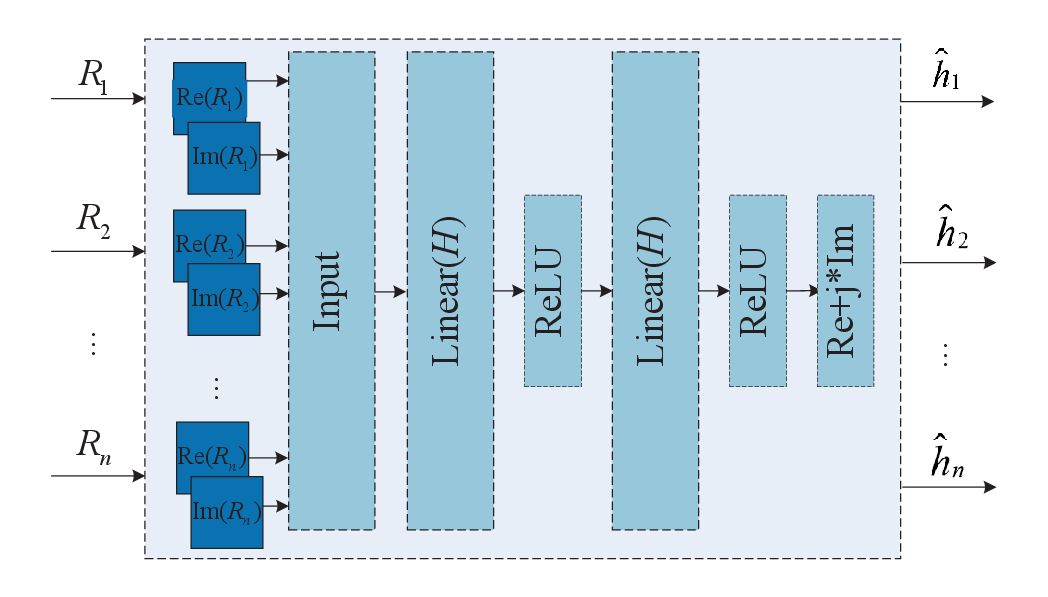}}
	\caption{Proposed MLP for FAS channel estimation}
	\label{fig:proposed}
\end{figure}

The resulting input dimension is $2PM$, and the output dimension is $2N$. The proposed network comprises three fully connected layers: the first two employ ReLU activation to capture nonlinear relationships, while the final layer is linear, producing the reconstructed channel vector. The hidden-layer operations can be expressed as
\begin{equation}
	\begin{aligned}
		\tilde{\mathbf{h}}
		&= f_{\boldsymbol{\Theta}}(\tilde{\mathbf{r}}) \\
		&= \mathbf{W}_3 \, \phi_2\!\left(
		\mathbf{W}_2 \, \phi_1\!\left(
		\mathbf{W}_1 \tilde{\mathbf{r}} + \mathbf{b}_1
		\right)
		+ \mathbf{b}_2
		\right)
		+ \mathbf{b}_3,
	\end{aligned}
\end{equation}
where $\mathbf{W}_i$ and $\mathbf{b}_i$ are the weight matrices and bias vectors of the $i$-th layer, respectively, and $\phi_i(\cdot)$ denotes the ReLU activation $\phi_i(x)=\max(0,x)$. A schematic of the overall MLP-based estimator is shown in Fig.~\ref{fig:proposed}.

\begin{itemize}
	\item[-]{\em 1) Data Preparation:}
	Under a fixed system configuration, a dataset $(\mathbf{R},\mathbf{H})$ is pre-generated or loaded from storage, where $\mathbf{R}\in\mathbb{R}^{PM\times N_{tr}}$ contains the received pilot signals and $\mathbf{H}\in\mathbb{R}^{N\times N_{tr}}$ contains the corresponding channel vectors generated from the channel model. The dataset is randomly split into a training set and a validation set with ratio $\rho$. To ensure stable training and avoid feature-scale imbalance, both $\mathbf{R}$ and $\mathbf{H}$ are normalized using standard score normalization:
	\begin{equation}
		\hat{\mathbf{R}}=\frac{\mathbf{r}-\mu_r}{\sigma_r}, 
		\quad
		\hat{\mathbf{H}}=\frac{\mathbf{h}-\mu_h}{\sigma_h},
	\end{equation}
	where $\mu$ and $\sigma$ are computed from the training set. All data are converted into 32-bit floating tensors and transferred to GPU when available.
	
	\item[-]{\em 2) MLP Training:}
	The proposed estimator is implemented as a fully connected feed-forward network parameterized by $\Theta$, consisting of two hidden layers, each with $H$ neurons followed by ReLU activation, and a linear output layer. This architecture balances nonlinear modeling power and computational simplicity. Formally, the network realizes
	\begin{equation}
		f_{\Theta} : \mathbb{R}^{\hat{\mathbf{R}}} \rightarrow \mathbb{R}^{\hat{\mathbf{H}}}.
	\end{equation}
	The model is trained using the Adam optimizer with an initial learning rate. Mini-batch training is adopted, and training samples are reshuffled before each epoch to enhance statistical robustness. After each epoch, validation loss is evaluated to monitor generalization performance.
	
	\item[-]{\em 3) Overfitting Prevention:}
	Early stopping is applied where the training terminates if the validation NMSE does not improve for 5 consecutive epochs, and the parameters achieving the lowest validation loss are retained. During testing, samples are normalized using training-set statistics, and the network outputs are inverse-transformed to recover the real and imaginary parts of the estimated channel vector.
\end{itemize}

Performance is quantified by the NMSE:
\begin{equation}
	\mathrm{NMSE}=\frac{\|\hat{\mathbf{h}}-\mathbf{h}\|_2^2}{\|\mathbf{h}\|_2^2},
\end{equation}
where $\hat{\mathbf{h}}$ and $\mathbf{h}$ denote the estimated and true channel vectors, respectively. The NMSE results are further converted to decibels. Convergence behavior is assessed through the evolution of training and validation NMSE curves across epochs.

\subsection{Complexity Analysis}
Let $D_{\mathrm{in}}$ and $D_{\mathrm{out}}$ denote the input and output dimensions, respectively, and let $H$ be the hidden-layer width. For a feed-forward neural network, computation is dominated by matrix–vector multiplications. For a single sample, the number of real multiplications during the forward pass is
\begin{equation}
	\mathcal{C}_{\text{forward}}
	= \mathcal{O}\!\left(D_{\mathrm{in}} H + H^2 + H D_{\mathrm{out}}\right).
\end{equation}
Backpropagation roughly doubles this cost:
\begin{equation}
	\mathcal{C}_{\text{back}}
	\approx 2\mathcal{C}_{\text{forward}}
	= \mathcal{O}\!\left(2H\left(D_{\mathrm{in}}+H+D_{\mathrm{out}}\right)\right).
\end{equation}

Let $N_{\mathrm{tr}}$ denote the number of training samples. Since each epoch processes all samples once, the per-epoch complexity is
\begin{equation}
	\mathcal{C}_{\text{per-epoch}}
	= \mathcal{O}\!\left(N_{\mathrm{tr}} \cdot 3H\left(D_{\mathrm{in}}+H+D_{\mathrm{out}}\right)\right).
\end{equation}
For $E$ training epochs, the overall training complexity is
\begin{equation}
	\mathcal{C}_{\text{all}}
	= \mathcal{O}\!\left(3EN_{\mathrm{tr}}H\left(D_{\mathrm{in}}+H+D_{\mathrm{out}}\right)\right).
\end{equation}

In the considered FAS channel estimation task, $D_{\mathrm{in}} = 2PM$ and $D_{\mathrm{out}} = 2N$. Therefore, the total training complexity becomes
\begin{equation}
	\mathcal{C}_{\text{all}}
	= \mathcal{O}\!\left(3 E N_{\mathrm{tr}} H \left(PM + H + 2N\right)\right).
\end{equation}
During online prediction, only a forward pass is needed, yielding per-sample complexity:
\begin{equation}
	\mathcal{C}_{\text{predict}}
	= \mathcal{O}\!\left(H\left(PM + H + 2N\right)\right).
\end{equation}

The computational complexities of the proposed method and existing schemes are summarized in Table~\ref{tab:complexity}.

\begin{table}[!t]
	\centering
	\caption{Computational Complexity of Different Schemes}
	\renewcommand{\arraystretch}{1.4}
	\begin{tabular}{r|l}
		\hline
		\textbf{Scheme}           & \textbf{Computational Complexity}                        \\ \hline
		FAS-OMP \cite{greedy1}                  & $\mathcal{O}\left(LPMN^2\right)$                         \\
		FAS-ML \cite{fas_beyes}                   & $\mathcal{O}\left(I_o PM L(PM+N)\right)$                 \\
		S-BAR (Stage 1) \cite{fas_beyes}           & $\mathcal{O}\left(PM\left(P^2M^2+NPM+N^2\right)\right)$  \\
		S-BAR (Stage 2)  \cite{fas_beyes}          & $\mathcal{O}(PMN)$                                       \\
		Proposed-Training Stage   & $\mathcal{O}\left(3E N_{\mathrm{tr}} H(2PM+H+2N)\right)$ \\
		Proposed-Predicting Stage & $\mathcal{O}\!\left(H(PM+H+2N)\right)$                   \\ \hline
	\end{tabular}
	\label{tab:complexity}
\end{table}
\begin{table}[!t]
	\caption{Simulation Parameters}
	\centering
	\renewcommand{\arraystretch}{1.2}
	\begin{tabular}{r|l}
		\hline
		\textbf{Parameter}                 & \textbf{Values}               \\ \hline
		Number of ports $N$                & 256                           \\
		Number of fluid antennas $M$       & 4                             \\
		Number of pilots $P$               & 64                            \\
		Carrier frequency $f_c$            & 3.5 GHz                       \\
		Antenna length $W$            & 10 wavelength                       \\
		Path gains                         & $\mathcal{C}\mathcal{N}(0,1)$ \\
		Incident angles                    & $\mathcal{U}(-\pi, +\pi)$     \\
		Max. angle spread                  & $5^{\circ}$                   \\
		Hidden layer size $H$              & 512                           \\
		Training samples $N_{\mathrm{tr}}$ & 5*$10^4$                        \\
		Validation ratio $\rho$            & 0.05                          \\
		Batch size                         & 256                           \\
		Learning rate                      & $10^{-4}$                     \\
		Early stopping patience            & 20                             \\ \hline
	\end{tabular}
	\label{tab:sim_params11}
\end{table}
In summary, the proposed MLP-based channel estimation method provides a computationally efficient alternative to traditional CS and AO-based approaches. The training complexity scales linearly with the number of samples and epochs, while online prediction remains lightweight due to the compact FCNN structure. This makes the proposed approach highly suitable for real-time FAS channel estimation in practical wireless communication systems.
\begin{figure}[!t]
	\centerline{\includegraphics[width=\columnwidth]{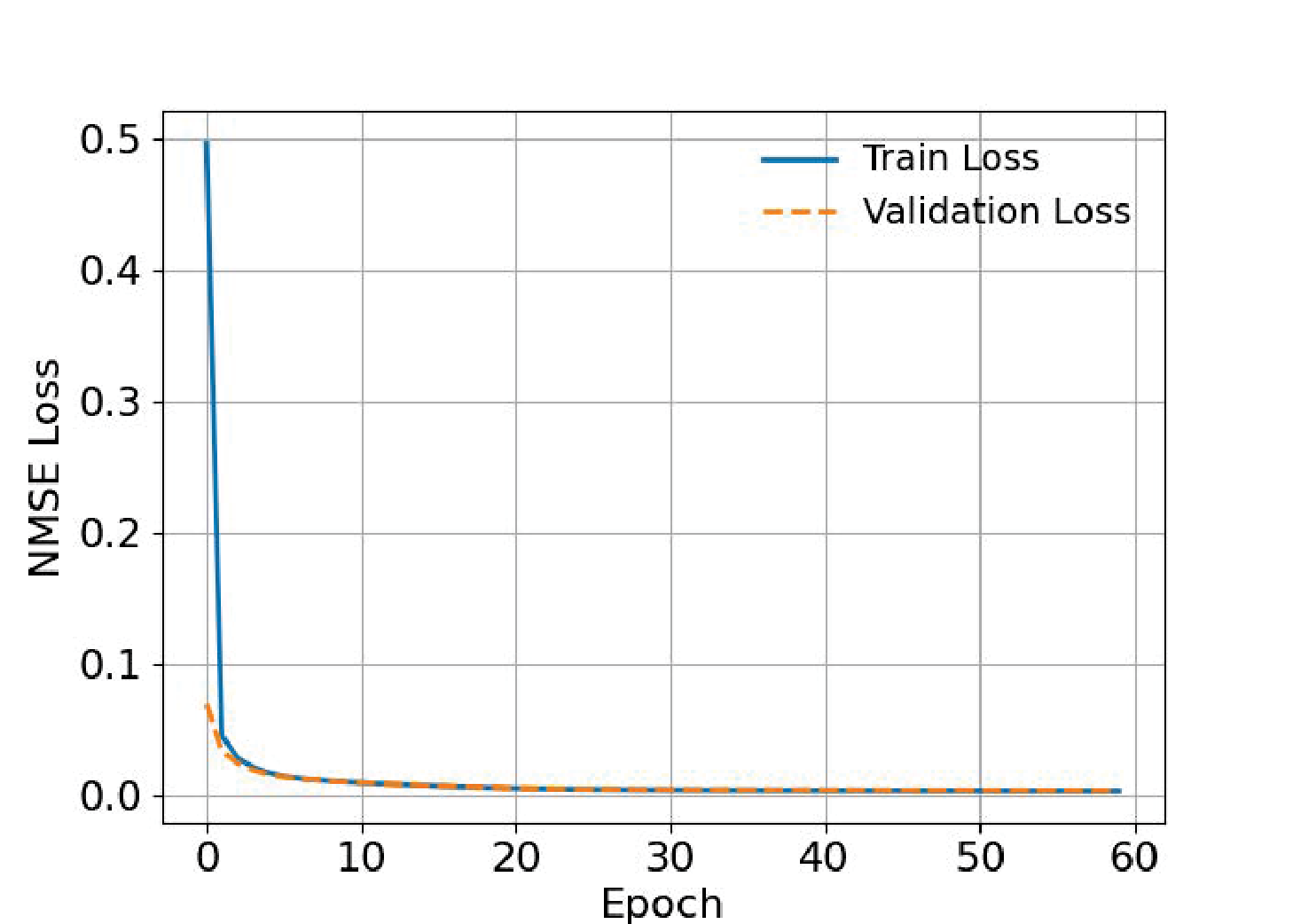}}
	\caption{Training and validation loss curves of the proposed network: Convergence behavior.}
	\label{fig:NMSE_convergence}
\end{figure}
\section{Simulation Results}\label{sec.numerical}
In this section, we present the simulation results to evaluate the performance of the proposed scheme. All experiments were conducted using Python with the PyTorch framework on a workstation equipped with an NVIDIA GPU. The default system parameters employed in the simulations are listed in Table~\ref{tab:sim_params11}.

Specifically, two representative configurations are considered to examine the impact of multipath richness: $(C, R) = (4,40)$ and $(C, R) = (2,10)$, where $C$ and $R$ denote the numbers of clusters and rays, respectively. {\em A small $(C,R)$, such as $(2,10)$, indicates a strongly sparse channel where only a few dominant scattering paths contribute to the received signal.} This typically corresponds to scenarios with limited scattering, e.g., suburban or rural environments, or mmWave links with highly directional propagation.  Conversely, {\em a large $(C,R)$, such as $(4,40)$, represents a weakly sparse or dense-scattering channel}. In such environments, the received signal consists of rich multipath components with significant angular spread, which is characteristic of indoor hotspots, urban microcells, or high-mobility rich-scattering environments.

The spatial sparsity level is set to $L = 2C$. The number of antenna ports is $N = 256$, the antenna length $W$ is fixed to 10 wavelength and the signal-to-noise ratio (SNR) is swept from $-15$ dB to $15$ dB to assess performance under different channel conditions. The proposed neural network is trained using the Adam optimizer with a learning rate of $10^{-4}$ and a batch size of 256. The obtained results are compared with several benchmark methods, including FAS-OMP \cite{greedy1}, FAS-ML \cite{fas_beyes}, SelMMSE \cite{ls_estimate}, and S-BAR \cite{fas_beyes}.

\begin{figure}[!t]
	\centerline{\includegraphics[width=\columnwidth]{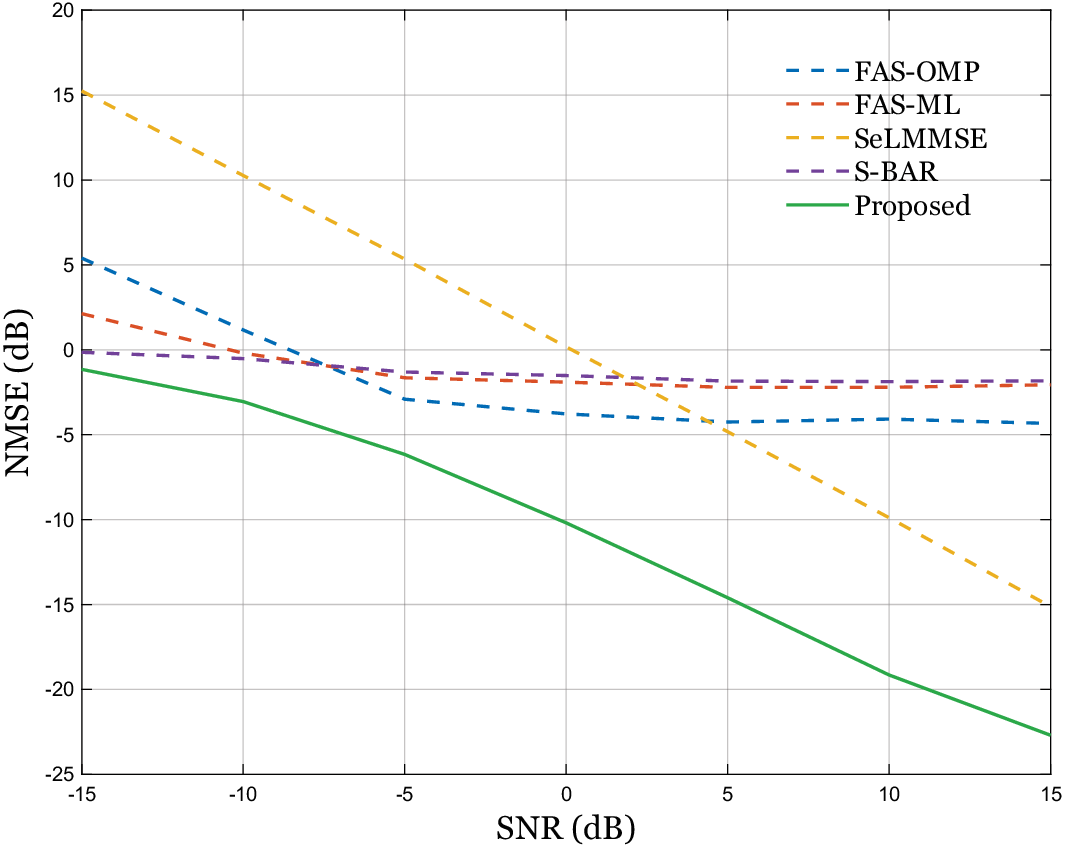}}
	\caption{NMSE performance under the strongly sparse CSC channel model with $(C, R) = (2,10)$. Benchmarks include the FAS-OMP \cite{greedy1}, FAS-ML \cite{fas_beyes}, SelMMSE \cite{ls_estimate}, and S-BAR \cite{fas_beyes}.}
	\label{fig:nmse_vs_snr1}
\end{figure}

\begin{figure}[!t]
	\centerline{\includegraphics[width=\columnwidth]{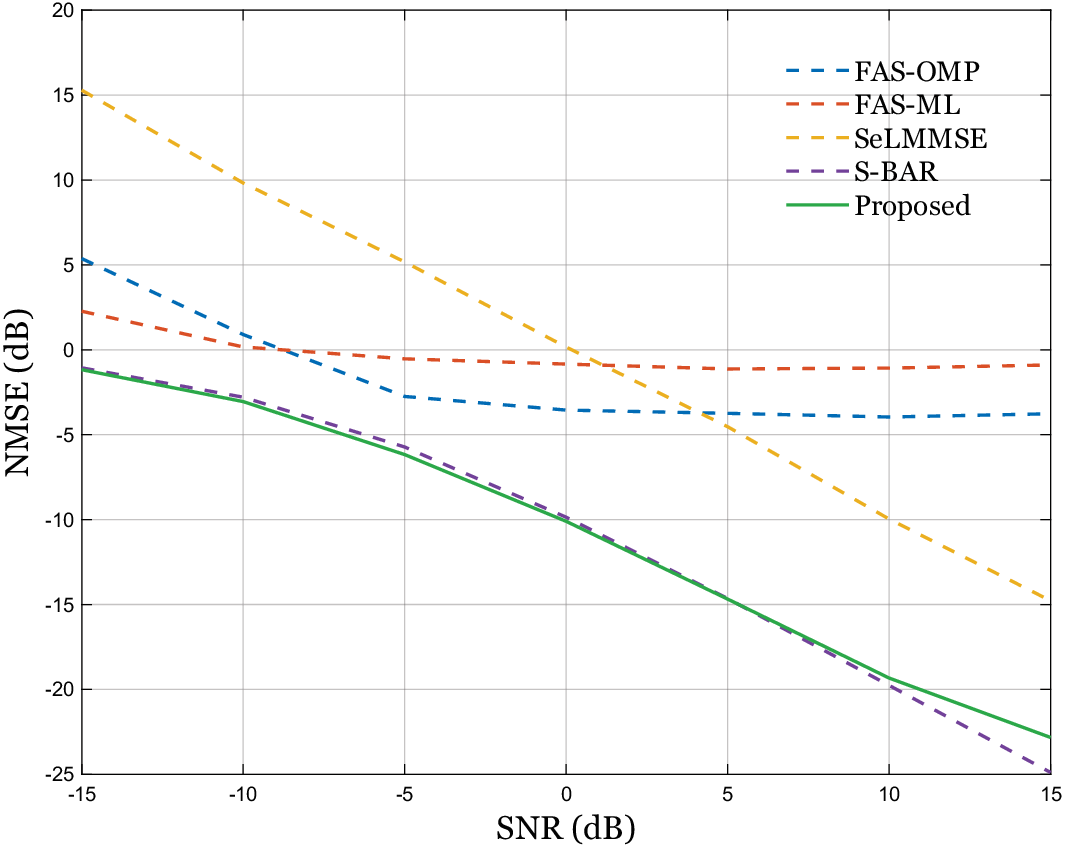}}
	\caption{NMSE performance under the the weakly sparse CSC channel model with $(C, R) = (4,40)$. Benchmarks include the FAS-OMP \cite{greedy1}, FAS-ML \cite{fas_beyes}, SelMMSE \cite{ls_estimate}, and S-BAR \cite{fas_beyes}.}
	\label{fig:nmse_vs_snr2}
\end{figure}
Fig.~\ref{fig:NMSE_convergence} illustrates the convergence behavior of the proposed neural network during training. Both the training and validation NMSE decrease rapidly within the first few epochs and gradually stabilize after approximately 30 epochs, demonstrating the effectiveness and stability of the learning process. The close alignment between the two curves indicates good generalization capability, with no observable overfitting. These results confirm that the adopted network architecture and training strategy facilitate efficient convergence and robust channel estimation across a wide range of SNR conditions.

Fig.~\ref{fig:nmse_vs_snr1} shows the NMSE performance of different channel estimation algorithms under the sparse multipath scenario $(C, R) = (2,10)$. The proposed method consistently outperforms all benchmark schemes, including FAS-OMP, FAS-ML, SelMMSE, and S-BAR, across the entire SNR range. The improvement becomes more pronounced in the high-SNR regime, where conventional approaches often suffer from noise sensitivity. For example, at $\mathrm{SNR} = 5~\mathrm{dB}$, the proposed estimator achieves an NMSE gain exceeding 10 dB relative to the second-best method. These results demonstrate the strong nonlinear modeling capability of the proposed network, enabling accurate channel reconstruction even under sparse scattering and low-SNR conditions.

Fig.~\ref{fig:nmse_vs_snr2} presents the results for the rich scattering scenario $(C, R) = (4,40)$. Similar to the previous case, the proposed model achieves lower NMSE than FAS-OMP, FAS-ML, and SelMMSE across all SNR levels. The performance gap between the proposed network and the S-BAR algorithm becomes small in this scenario, as S-BAR is able to exploit well-trained empirical kernels to effectively capture the underlying channel characteristics in rich multipath conditions. Nevertheless, the proposed network maintains competitive estimation accuracy while offering advantages in convergence speed and robustness. These results confirm that the proposed method generalizes well across different scattering densities and provides reliable channel reconstruction in diverse propagation environments.
\section{Conclusion}\label{sec.conclusion}
In this paper, a deep learning-based channel estimation framework is proposed for the CSC channel model, where the inherent nonlinear mapping between the received signal and the true channel response is efficiently learned through a MLP network. Simulation results demonstrate that the proposed method achieves significantly lower NMSE compared to conventional estimators such as FAS-OMP, FAS-ML, SeLMMSE, and S-BAR, under both sparse and rich scattering conditions. The superiority of the proposed model is particularly evident in low-SNR scenarios, highlighting its robustness to noise and modeling inaccuracies. Furthermore, the network shows strong generalization capability across various channel configurations, confirming its potential for real-time and adaptive channel reconstruction in practical wireless communication systems. Future work will focus on extending the proposed framework to dynamic FAS environments and exploring hybrid model-driven learning architectures to further enhance performance and interpretability.
\balance

\end{document}